\documentclass[11pt, conference, compsocconf, onecolumn]{IEEEtran}
\usepackage{graphicx}

\begin{document}
\title{Analyzing Delay in Wireless Multi-hop Heterogeneous Body Area Networks}

\author{N. Javaid, M. Yaqoob, M. Y. Khan, M. A. Khan, A. Javaid, Z. A. Khan$^{\$}$\\
        COMSATS Institute of Information Technology, Islamabad, Pakistan. \\
        $^{\$}$Faculty of Engineering, Dalhousie University, Halifax, Canada.\\
     }

\maketitle

\begin{abstract}
With increase in ageing population, health care market keeps growing. There is a need for monitoring of health issues. Wireless Body Area Network (WBAN) consists of wireless sensors attached on or inside human body for monitoring vital health related problems e.g, Electro Cardiogram (ECG), Electro Encephalogram (EEG), ElectronyStagmography (ENG) etc. Due to life threatening  situations, timely sending of data is essential. For data to reach health care center, there must be a proper way of sending data through reliable connection and with minimum delay. In this paper transmission delay of different paths, through which data is sent from sensor to health care center over heterogeneous multi-hop wireless channel is analyzed. Data of medical related diseases is sent through three different paths. In all three paths, data from sensors first reaches ZigBee, which is the common link in all three paths. Wireless Local Area Network (WLAN), Worldwide Interoperability for Microwave Access (WiMAX), Universal Mobile Telecommunication System (UMTS) are connected with ZigBee. Each network (WLAN, WiMAX, UMTS) is setup according to environmental conditions, suitability of device and availability of structure for that device. Data from these networks is sent to IP-Cloud, which is further connected to health care center. Delay of data reaching each device is calculated and represented graphically. Main aim of this paper is to calculate delay of each link in each path over multi-hop wireless channel.
\end{abstract}

\section{Introduction}
Many countries face ageing population, as number of senior citizens increasing all over the world. With increase in ageing population, there is a need to monitor their health on regular basis. Specialized health monitoring of serious cases are very important, however, it is quiet expensive. With emerging technology, remote patient monitoring is possible. WBAN consists of interconnected sensors, either placed around body or small enough to be placed inside the body. It provides ease of connectivity with other systems and networks thus allowing proper health monitoring. With help of WBAN, monitoring of patient is done remotely through internet, intranet or any other network. These sensors continuously monitor data and send it to health care center.\\
\indent In patient monitoring system data transmission reliability with low delay is very important. Different technologies have been used in sending of medical data to health care center like bluetooth connected to cellular system[1]. In this paper, we present three different paths, which can be employed at different places for the monitoring of serious critical data of ECG, EEG etc. Different devices connected with WBAN takes data from sensor nodes and transmits it to health care center.\\
\indent A low cost and reliable approach is preferred for transmission of data from sensor nodes to health monitoring room. Heterogeneous multi-hop transmission represents sending of medical data to health monitoring room using multiple devices that are interconnected with each other, each performing its own operation. In this paper we implement three different paths for sending of data. Each path is implemented according to need and environmental conditions. We calculate delay of all links that takes part in data sending for all three paths. ZigBee network is connected to nodes attached to body, for better monitoring and control.\\
\indent Bluetooth considered to be a network that has easy connection with cellular systems. Both ZigBee and Bluetooth offer low power consumption, respective data rate and both are useful for short range. However, potential interference from different devices is a concern for Bluetooth[1]. ZigBee consumes less power than Bluetooth. WLAN is connected with ZigBee, which is referred as path 1 for data transmission. It consists of 802.15 has easy connectivity and supports Wireless Fidelity (Wi-Fi). WiMAX is connected with ZigBee in path 2, WiMAX provides high data rate with long coverage area. Path 3 comprises of ZigBee connected with UMTS.\\
\indent In this work, we focus on calculating delays of all three paths, by inspecting delays of each link of each path. We consider three hop structure for each path, which consists of a ZigBee link from sensor nodes to ZigBee device, second hop from ZigBee device to WLAN, WiMAX or UMTS and third hop from WLAN, WiMAX or UMTS to IP-Cloud as shown in fig 1.\\
\begin{table*}
\caption{Properties of ZigBee, WLAN, WiMAX and UMTS}
\begin{center}
    \begin{tabular}{| p{1.3cm} | p{2.5cm} | p{2cm} | p{2cm} | p{1.3cm} | p{1.3cm}| p{1.5cm}| p{1.4cm}| p{1.5cm}|}
    \hline
    Technology & Advantages & Disadvantages & Application & Network & Modulation & Network Connectivity & Network Topology&Access Protocol \\ \hline
    ZigBee & Low cost, Ultra low power consumption, easy installation & Short range, Low data rate & Sensor networks & Mesh&DSSS & Yes & Ad-Hoc & CSMA/CA \\ \hline
    WLAN & Flexibility and Robustness  & Low QoS & Wi-Fi, Internet & IP, P2P & DSSS & No & Infrastructure & CSMA/CA  \\ \hline
    WiMAX&High Data Rate, Faster Deployment&LOS required, Weather affects&Broadband Internet Connectivity &IP&QAM&Yes&Infrastructure&Request/Grant\\ \hline
    UMTS & High Data Rate& Handover problems & Wi-Fi, Bluetooth, Infrared &Cellular/3G&DQPSK&Yes&Ad-Hoc& WCDMA\\ \hline
    \end{tabular}
\end{center}
\end{table*}
\indent Properties of ZigBee, WLAN, WiMAX and UMTS are shown in table I. ZigBee is a low cost device having low power consumption. It is based on Carrier Sense Multiple Access with Collision Avoidance (CSMA/CA). It uses modulation scheme of Direct Sequence Spread Spectrum (DSSS).\\
\indent WLAN has low Quality of Service (QoS). It uses Direct Sequence Spread Spectrum (DSSS) modulation scheme and supports infrastructure based topology. It has Internet Protocol (IP) and Peer to Peer network (P2P).\\
\indent WiMAX has high data rates as compared to WLAN and ZigBee with high coverage range. Line of Sight (LOS) is basic requirement while using WiMAX. Drawback of using WiMAX is that, it is affected by changing weather conditions. It request for access protocol or it is granted by network by default. Quadrature Amplitude Modulation (QAM) is the modulation scheme is used in WiMAX.\\
\indent UMTS supports high data rate, however due to cellular structure, handover problems exist in it, due to which communication is affected. It has Ad-hoc network topology and Wide Code Division Multiple Access protocol (WCDMA). Each network has different applications based on their structure and type of scheme they use.\\
\section{Related Work and Motivation}
\indent Authors in \cite{9} propose a MAC protocol for WBAN using wakeup radio mechanism. TDMA based scheme combined with wakeup radio is used to design an energy efficient Medium Access Control (MAC). However, their simulations are only for single hop communication. Moreover authors have not done calculations for heterogeneous network environment, as we do it in our paper.
\\
\indent In paper \cite{3}, authors adopt a tree protocol for ECG signal carried over ZigBee. A prototype of DSP platform enabling good performance in ZigBee is established. Symmetrical system architecture is developed. They develop mathematical model and simulated transmission time delay of ECG data. Mathematical model is built for CSMA/CA mechanism. However authors consider ZigBee. As ZigBee is for low distance coverage and low data rate, we consider a heterogeneous network environment for sending of medical data through WLAN, WiMAX and UMTS networks.
\\
\indent WBAN is used to develop a patient monitoring system which offers flexibility and mobility to patients. It allows flexibility of using remote monitoring system via either internet or intranet. Performance of IEEE 802.15.4/ZigBee operating in different patient monitoring environment is examined in \cite{4}. However, authors simulate hospital network based on Ethernet standard. Also authors not discussed mathematical aspects of their environment setup, which we do in our paper.
\\
\indent Authors present a new cross-layer communication protocol for WBANs,  CICADA in \cite{7}. This protocol setup a network tree in a distributed manner. This structure is used to guarantee collision free access to the medium and to route data towards the sink. Energy consumption is low because nodes can sleep in slots, where they are not transmitting or receiving. However, their propose protocol only support node to sink traffic.
\\
\indent Authors in paper \cite{2}, state that, IEEE 802.15.4 standard is designed as a low power and low data rate protocol with high reliability. They analyze unslotted version of protocol with maximum throughput and minimum delay. The main purpose of IEEE 802.15.4 standard is to provide low power, low cost and highly reliable protocol. The standard defines a physical layer and a MAC sub layer. This standard operates in either beacon enabled or non beacon mode. Physical layer specifies three different frequency ranges, 2.4 GHz band with 16 channels, 915 MHz with 10 channels and 868 MHz with 1 channel. Calculations are done by considering only beacon-enabled mode and with only one sender and receiver. However, it has high cost of power consumption. As number of sender increases, efficiency of 802.15.4 decreases. Throughput of 802.15.4 declines and delay increases when multiple radios are used because of increase in number of collisions.
\\
\indent Authors in \cite{6}, examine use of IEEE 802.15.4 standard in ECG monitoring sensor network and study the effects of CSMA/CA mechanism. They analyze performance of networks in terms of transmission delay, end to end latency, and packet delivery rate. For time critical applications, a payload size between 40 and 60 bytes is selected due to lower end to end latency and acceptable packet delivery rate. However, authors consider only single hop communication. We calculate and implement it with heterogeneous networks multi-hop transmission.
\\
\indent In paper \cite{10}, authors discuss three emerging wireless standards, ZigBee, WiMAX and Wi-Fi. They are discussing in this paper, characteristics and application of these emerging technologies. Comparisons are presented to prove which performs better. However, no scenario is considered, in which all these technologies are implemented or combined with one another and nothing is proved graphically.
\\
Authors evaluate overall transmission delay of ECG packets over two-hop wireless channel.  ECG data from  BAN are compressed and sent through a Bluetooth-enabled ECG monitor to a smart phone and thereafter to a cellular Base Station (BS). Exploiting the inherent heartbeat pattern in ECG traffic, they introduce a context aware packetization for ECG transmission in \cite{1}. However, ECG is not only data that can be transmitted, other medical data through different networks is sent. In this paper we calculate delay of heterogeneous networks of multi-hop which is not done in this paper.
\\
\indent A Traffic-adaptive MAC protocol (TaMAC) is introduced by using traffic information of sensor nodes in \cite{8}. TaMAC protocol is supported by a wakeup radio which is used to support emergency and on-demand events in a reliable manner. Authors compare TaMAC with beacon-enabled IEEE 802.15.4 MAC, WiseMAC, and SMAC protocols. They study co-existence of heterogeneous WBAN traffic. However, authors have not simulated delay in a heterogeneous environment.
\\
\indent An application of wireless cellular technologies CDMA2000 1xEVDO, as a promising solution to wireless medical system is propose in \cite{5}. Authors analyze end-to-end delay analysis for medical application using CDMA2000 1xEVDO. However, they only consider worst-case end to end delay over cellular network. They have not discussed about interoperability of CDMA2000 1xEVDO with WBAN. Also they only analyze mathematical equations for ECG data. In our paper, we analyze for general medical traffic.
\section{Network Architecture for Heterogeneous Multi-hop Network in WBAN}
One of most important metrics for QoS is delay. There is a need to know delay of data passing through different paths and devices. It is highly dependent on type of structure used, protocols used in networks and requires high level of reliability. In this paper we calculate delay of all links that takes part in sending data from different device to the health center.
\begin{eqnarray}
D_{total}= D1+D2+D3
\end{eqnarray}
$D_{total}= Total$ $delay$ $from$ $sensor$ $node$ $to$ $server$
\\
$D1=Delay$ $of$ $link 1$
\\
$D2=Delay$ $of$ $link 2$
\\
$D3=Delay$ $of$ $link 3$
\\
\begin{figure}[!h]
\centering
\caption{Heterogeneous Multi-hop Network in WBAN}
\includegraphics[width=3.5 in, height=2.5 in]{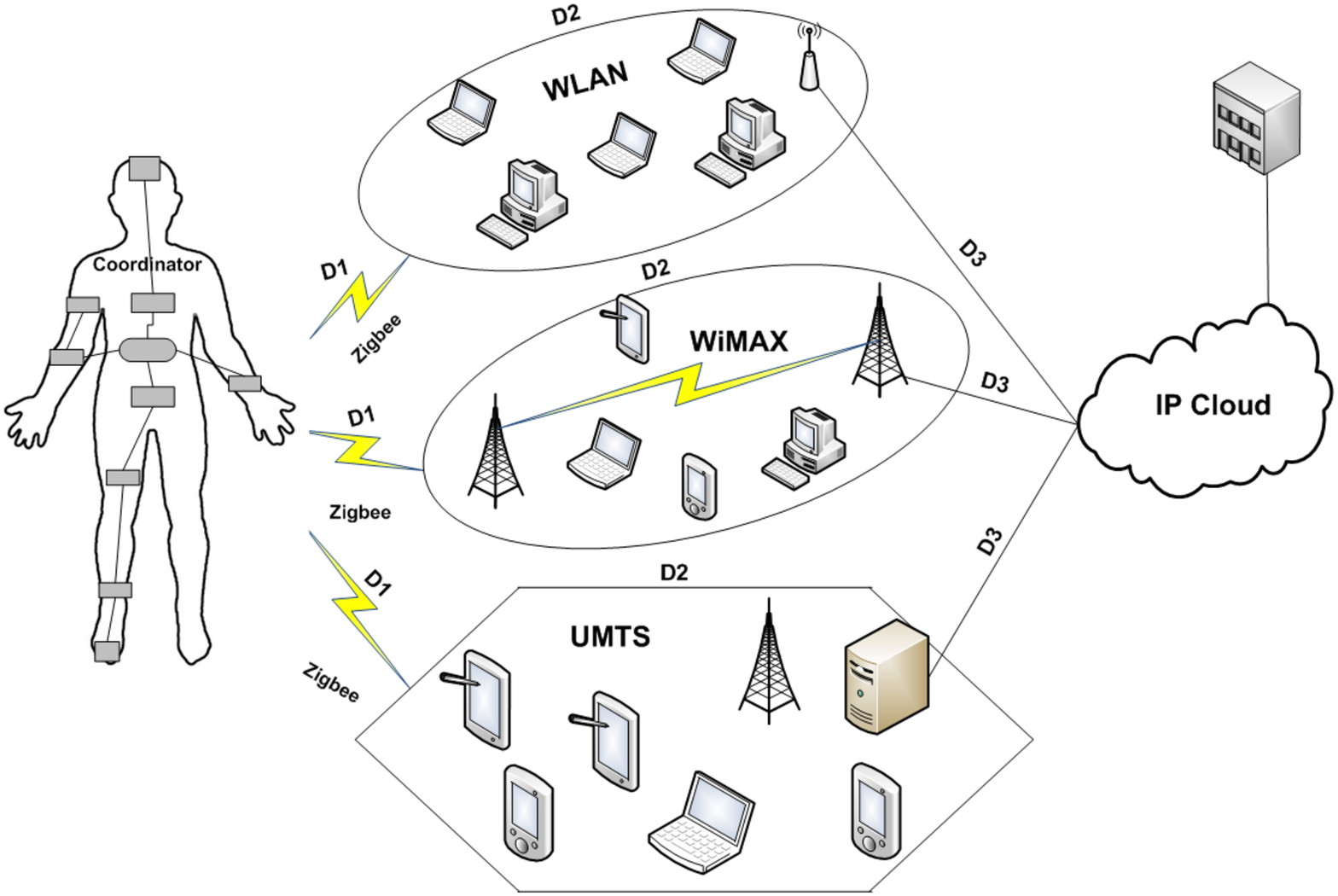}
\end{figure}
\indent Fig 1 shows complete structure of data transmission from three different paths to health monitoring room. Total delay is calculated by equation 1. Delays of each link are shown in fig 1. Delay 1 (D1) is delay from nodes connected with human body to ZigBee, Delay 2 (D2) is from ZigBee to ext connected network WLAN, WiMAX or UMTS and Delay 3 (D3) is from WLAN, WiMAX or UMTS to IP-cloud. \\
\indent In first path data travels from nodes attached to human body  towards ZigBee as shown in fig 1. After passing on to ZigBee, data divides according to path. If first path is established then data goes towards WLAN and devices attached to WLAN. If second path is implemented then data passes on to WiMAX. If path 3 is established then data passes on to UMTS.\\
\indent Delay of all these paths are referred as delay 2 according to the implementation shown in fig 1. After data is passed from any of the three networks WLAN, WiMAX, UMTS, it is send to IP-cloud which is connected to health care center.\\
\indent All technologies are implemented according to need, ZigBee is used as starting device, because it is best short range network. After ZigBee, if we want to develop a Wi-Fi structure for sending of data to the monitoring room then, best device is WLAN which supports Wi-Fi internet with reliability and gives good coverage area. If we want to develop a last mile network and also have well developed structure of WiMAX than it is the network that is needed to be used. It supports high data rate and also gives last mile connectivity. If want to develop a cellular structure based on mobile traffic then after ZigBee, UMTS structure is established, which is based on cellular structure and supports all kind of mobile traffic with good coverage area. Handoff is a problem in UMTS, but it is minimized by using different techniques.
\subsection{UMTS}
It is based on GSM and relates to 3rd generation of wireless technology. It uses Wideband Code Division Multiple Access (W-CDMA) to provide greater efficiency and bandwidth to mobile network operators. UMTS supports maximum data transfer rates up to 42Mbit/s.
\subsection{WLAN}
It links more than one device by using some wireless method. It connects through an access point to internet. This gives mobility to move around and still connected to internet. WLAN is based on IEEE 802.11 like ZigBee. It is most popular end user device due to cheap cost, ease of installation and wireless access to users.
\\
\indent Many projects are setup using WLAN to provide wireless access to different locations. WLAN supports data rate up to 54Mbit/s. It is long term and cost effective. It is easier to add more stations with WLAN. It provides connectivity in those areas where there are no cables.
\subsection{ZigBee}
It is small, low-power digital radio based on IEEE 802.15.4 standard for Personal Area Networks. It is less expensive than other Wireless Personal Area Networks (WPANs), such as Bluetooth. ZigBee is used in those application where we need low data rate, long battery life and secure networking. It supports data rate up to 250 Kbit/s. It is best suited for periodic data or single signal transmission from sensor to other device input. Low cost of ZigBee allows it to be widely develop in wireless control and monitoring applications like sending of medical data from sensor node on human body to other devices. Its network layer supports both star, tree and mesh networks. ZigBee builds up MAC layer and Physical layer for low data rate WPANs.
\\
\indent Nodes in ZigBee go from sleep to active mode in 30ms or less due to which its latency is low. Due to long sleep time of nodes, its power consumption is low and gives long battery life. Its protocols are intended for low data rates and have low power consumption thus resulting network uses small amount of power. ZigBee devices are of three types:
\\
\indent ZigBee Coordinator (ZC): This is root of all network and makes bridges with other networks. There is exactly one ZigBee coordinator in each network because it starts the network.
\\
\indent ZigBee Router (ZR): It runs an application function and can act as router passing data from other devices.
\\
\indent ZigBee End Device (ZED): It only communicates with the coordinator or router. Nodes in ZED are in sleep mode thereby giving long battery life. ZED requires least amount of memory and therefor it is less expensive than ZR and ZC.
\subsection{WiMAX}
It is a standard made for wireless communication providing 120 Mbit/s data. It easily passes range of WLAN, offering area network with signal radius of 50km. It provides data transfer rates that are superior to cable-modem and Digital Subscriber Line (DSL) connections. It is based on IEEE 802.16 family of networks standards. WiMAX is similar to Wi-Fi but provides much greater data rate and greater distance.
\\
\indent WiMAX internet connectivity is provided to multiple devices and it is connected to multiple devices to provide internet to home, business and other places. Standard 802.16 is versatile enough to accommodate Time Division Duplexing (TDD) and Frequency Division Duplexing (FDD).
\section{Heterogeneous Multi-hop Paths}
In all three parts, sensor node collects medical data from body and sends it to ZigBee, ZigBee being the first device connected to human body sensors in all three paths. In path 1 sensor nodes, ZigBee, WLAN, IP-cloud and Server are connected. Path 2 consists of sensor nodes, ZigBee, WiMAX, IP-cloud and health monitoring room. Path 3 consists of sensor nodes, ZigBee, UMTS, IP-cloud and health monitoring room. Sensor nodes are attached to body of a person, from where they record the medical data and sends it to ZigBee. ZigBee through end device, receives data and passes it to coordinator, so that, coordinator passes data to router from where data is sent to WLAN in path 1 and to WiMAX in path 2. Parameters are set according to the requirement. Packet Inter arrival Time is 0.04 seconds with packet size of 1024 bytes.
\subsection{Structure Components of all devices}
\indent Structure of WLAN consists of application layer, data link layer, transport layer and physical layer. Data from all these layers is processed and then passed to IP-cloud. Data when reaches WiMAX, it passes through WiMAX user station, IP backbone and WiMAX base station (BS). As ZigBee operates on 802.15.4 hence it has CSMA/CA mechanism. Minimum backoff exponent is kept to 2, with maximum number of backoff to 3. Channel sensing duration is 0.1 seconds. Network parameters that are used in ZigBee coordinator and end device are given in table II.
\begin{table}
\caption {Network Parameters Of ZigBee coordinator and End device}
\begin {center}
\begin {tabular} {| p{4cm} | p{2cm} |}
\hline
Parameters type & Value \\ \hline
Beacon Order &   6 \\ \hline
Superframe Order & 0 \\ \hline
Maximum Routers & 5 \\ \hline
Maximum Depth & 5 \\ \hline
Beacon Enabled Network & Disabled \\ \hline
Mesh Routing  & Disabled \\ \hline
Route Discovery Timeout  & 10(msec) \\ \hline
Back off Exponent & 3 \\ \hline
Maximum Number Of Backoffs & 2\\ \hline
\end{tabular}
\end{center}
\end{table}
\\
\indent In this table beacon order of IEEE 802.15.4 is set to 6. When we look at structure of IEEE 802.15.4 we know about the beacon order[3]. Superframe order has also been explained in paper [3]. Maximum number of routers that can be connected to coordinator and end device is kept to 5 with each having depth of 5. Route discovery timeout is kept to $10\mu$sec.
\subsection{Structure of WLAN}
\indent In case of WLAN there are many paraments that needed to be set before start of communication between WLAN and ZigBee, routing protocols in it are kept to default so that routing takes place as normally. IP is kept to constant, so that WLAN can easily communicate. There are two WLAN MAC addresses that need to be set for this communication one is $IF0_{P0}$ and other one is $IF1_{PO}$. Both these addresses are of different LAN parameters that needed to be set. Parameters of these MAC addresses are given in table III.\\
\begin{table}
\caption {WLAN PARAMETERS OF $IF0_{P0}$ and $IF1_{P0}$}
\begin {center}
\begin {tabular} {| p{4cm} | p{2cm} |}
\hline
Name of Parameter & Value \\ \hline
BSS Identifier & Auto Assigned \\ \hline
Physical technique &   Direct Sequence \\ \hline
Data Rate & 11Mbps \\ \hline
Transmit Power & 0.005 \\ \hline
Packet Reception threshold power & -95 dBm \\ \hline
Buffer size & 25600 bits \\ \hline
Large Packet Processing & Drop \\ \hline
\end{tabular}
\end{center}
\end{table}
\indent In this table Basic Service Set Identifier (BSSID) is the MAC address of station in an access point. It is administrated MAC address generated from 42 bit address. It is either kept to auto assigned so that WLAN assigns different BSSID to different stations or it is assigned by user. Each BSSID of device is separated from other, so that, conflict of same BSSID does not occur. Physical technique that is used in CSMA/CA is DSSS, input data is encrypted with a unique signal and then at the output it is decrypted using same code and data is recovered [3]. Data rate has been kept to 11Mbps for this device. Minimum power at which a node accepts a packet is referred as packet reception threshold power. Sending of large packet are set to drop, if there is a large packet waiting to be send, then it is dropped because large packet cause more delay and in this critical system we need packet with less delay.
\subsection{Structure of WiMAX}
\indent Parameter values in WiMAX are adjusted according to requirement. MAC address is auto assigned and IP routing protocols are kept to default. WiMAX contains base station (BS) parameters which include Code Division Multiple Access (CDMA), are kept to 8. Rest of the parameters with their values are given in table IV.
\begin{table}
\caption {WiMAX Parameters Value}
\begin {center}
\begin {tabular} {| p{4cm} | p{2cm} |}
\hline
Name of Parameter & Value \\ \hline
Antenna Gain & 15 dBi \\ \hline
Number of Transmitters & SISO \\ \hline
MAC Address &   Auto Assigned \\ \hline
Maximum Transmission Power  & 0.5W \\ \hline
Physical profile & OFDM 20MHz \\ \hline
Maximum number of SS nodes & 100 \\ \hline
Minimum Power Density & -110(dBm) \\ \hline
Maximum Power Density & -60(dBm) \\ \hline
\end{tabular}
\end{center}
\end{table}
\\
\indent In table IV antenna gain of WiMAX is set to 15 dBi. MAC addresses are kept to auto assigned WiMAX auto assigns the MAC address of devices to avoid conflict of addresses. Transmission power that is be transmitted by antenna is 0.5W. Physical Profile which is used by WiMAX is OFDM at 20MHz of frequency.
\subsection{Structure of UMTS}
\indent In path three sensor nodes send packets containing information about health to ZigBee and from ZigBee it is send to UMTS from where it is transferred to IP-cloud and server. ZigBee performs similar to the two path explained above and settings of parameters are also same. UMTS is a universal mobile telecommunication system which works on cellular technology. ZigBee is allowed to send data to UMTS by changing its inner structure, UMTS receives data from the end device of ZigBee, working as a router through Node B, than data is passed through Authentication Authorization and Accounting (AAA) server where authentication take place. Authorization confirms subscribers configuration information and finally collection billing information.  After all this process data is passed to UMTS Node B.
\\
\indent UMTS structure contains Home Location Register (HLR). HLR is centralized database unit that contains all data of mobile phone connected to core network. It also stores data of every Subscriber Identification Number (SIM) card issued by companies to customers. It contains UMTS Serving GPRS Support Node (SGSN) which works as Mobile Switching Center. UMTS Radio Network Controller (RNC) which works as Base Station Controller (BSC) and UMTS Base Station (BS). Data from networks passes to the UMTS server connected to router which is connected to IP-Cloud after passing through BS, BSC, RNC, SGSN and HLR. Flowchart in fig 2 shows complete structure of different devices combining in sending of data.\\
\indent Server of UMTS contains application, transport, routing and  data link layer in its structure each performs its own operation and these layers are set according to use. UMTS HLR, RNC and Router consists of complex inner structure. UMTS parameter values are given in table V.
\begin{figure}[!b]
\centering
\caption{Flow Diagram of Heterogeneous Multi-hop environment}
\includegraphics[width=4 in, height=5.5 in]{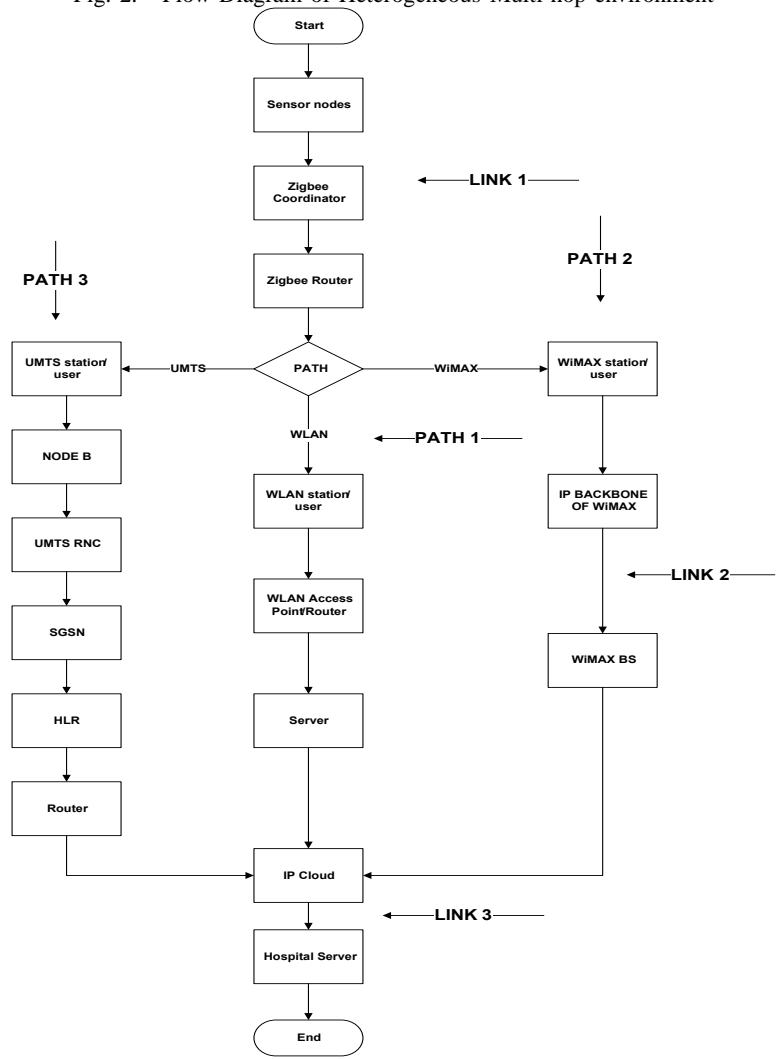}
\end{figure}

\begin{table}
\caption {UMTS Parameters value}
\begin {center}
\begin {tabular} {| p{3.5cm} | p{2cm} |}
\hline
Name of Parameter & Value \\ \hline
Processing time & 0.002 (sec) \\ \hline
Maximum retry on time expiry & 4 \\ \hline
Cell path loss parameters & Default \\ \hline
UMTS cell id & Default \\ \hline
UMTS SGSN ID & 0 \\ \hline
IP & Default \\ \hline
\end{tabular}
\end{center}
\end{table}
\section{Mathematical Modeling}
Link one in all three paths and link two in path 1 works on CSMA/CA mechanism, so transmission delay in it is calculated by the equation 2. T is the total transmission time that is needed to send data from sender to receiver. Backoff time is kept to 3. Backoff time may vary up to 5 depending on the parameters of the network device and traffic load. Inter Frame Space is delay that comes after any type of communication in CSMA/CA. Turnaround time is the transmission time of sending a packet and then receiving acknowledgement successfully[3].
\begin{eqnarray}
D=T_{bo}+T_{data}+T_{ta}+T_{ack}+T_{ifs}
\end{eqnarray}

Data transmission time $T_{data}$, Backoff slots time $T_{bo}$, Acknowledgement time $T_{ack}$ are given by equation 2, 3, 4 respectively[3].
\\
\begin{eqnarray}
T_{data}=\frac{L_{phy} + L_{mac hdr} + payload + L_{mac ftr}}{R_{data}}\\
\nonumber\\
T_{bo}=bo_{slots} * T_{boslots}      \\
\nonumber\\
T_{ack}= \frac{L_{phy} + L_{mac hdr} + L_{mac ftr}}{R_{data}}
\end{eqnarray}
\\
Where
\\
T$_{bo}$=Back Off Period
\\
T$_{data}$=Transmission Time of Data
\\
T$_{ta}$=Turn Around Time
\\
T$_{ack}$=Acknowledgement Transmission Time
\\
T$_{ifs}$=Inter Frame Space
\\
T$_{phy}$=Length of Physical header
\\
L$_{mac hdr}$=Number of MAC header
\\
Payload=Number of data byte in packet
\\
L$_{mac ftr}$=size of MAC footer
\\
$bo_{slots}$=Number of back off slots
\\
$T_{bo slots}$= Time for a back off slot
\\\\
\indent In CSMA/CA mechanism packet losses due to collision. Collision occurs, when more than one node, sensing medium and finding it idle at same time. If acknowledgement time is not taken in to account then there is no retransmission of packet and it is considered that each packet is delivered successfully.
The probability of end device successfully transmitting a packet is modeled as follows[3].
\\
\begin{eqnarray}
P_{backoff period}=\frac{1}{2^{BE}}
\\
\nonumber\\
P_{tss}= \frac{1}{D}{(1-\frac{1}{D})}^{BE-2} \\
= p({1-p})^{BE-2} \nonumber
\end{eqnarray}
\indent Where D is the number of end devices that is connected to the router or coordinator. BE is the backoff exponent in our case it is 3. $P_{tss}$ is the probability of transmission success at a slot. $\frac{1}{D}$ is the probability of end device successfully allocated a wireless channel.\\
\indent General formula for $P_{time delay event}$ is given by equation 8. Probability of time delay caused by CSMA/CA backoff exponent is estimated as in [2]. Value of BE is used in estimation. Minimum value of BE is 2 and we estimate it by applying summation from 2 to 3. $P_{tde}$ is the probability of time delay event.
\\
\begin{eqnarray}
P_{tde}= \sum_{n=0}^{2^{BE-1}} n\frac{1}{2_{BE}}p{1-p}^{BE-2}
\end{eqnarray}
\begin{eqnarray}
P_{tde}= \sum_{n=0}^{3} n\frac{1}{2_{BE}}p + \sum _{n=4}^{11} n\frac{1}{2_{BE}}p
\end{eqnarray}
\indent Expectation of the time delay is obtained as follows[3]. $P{E_{A}}$ and $P{E_{B}}$ are taken from equations 8 and 9 respectively.
\\
\begin{eqnarray}
E[Time Delay]=P{E_{A}|E_{B}}\\
\nonumber\\
= \frac{\sum_{n=0}^{3} n\frac{1}{2_{BE}}p + \sum _{n=4}^{11} n\frac{1}{2_{BE}}p}{\sum_{n=0}^{2^{BE-1}} n\frac{1}{2_{BE}}p{1-p}^{BE-2}}
\end{eqnarray}
\section{Simulations and Results}
OPNET Modeler is used for simulations and modeling. Simulations are ran for 1 hour, update interval is kept to 5000 events. Configure/Run values are given in table VI. Individual statistics are marked so that, their graphs are plotted. Individual statistics are of two types Global Statistics (GS) and Node Statistics (NS). GS is the measurement of statistics globally through overall structure and NS is the statistics observing at each node.
\subsection{Path 1}
In this path data passes through ZigBee router, WLAN and IP-Cloud to reach receiver from nodes. Parameter values for this path are given in table II. GS and NS that are marked for graphical data of both ZigBee and WLAN are as follows. Graphs have been plotted in stacked statistics and averaging time.
\\
\indent GS of WLAN and ZigBee : Delay (sec), Medium Access Delay(MAD)(sec)\\
\indent NS of WLAN and ZigBee : Delay (sec), Medium Access Delay(sec)
\\
\indent Delay in GS and NS of WLAN represents the end to end delay of all the packets received by the WLAN MACs of all the WLAN nodes in network and forward it to the higher layer. MAD represents the statistics for total queuing and contention delays of data, management, delayed block ACK, and block ACK request.\\
\indent Delay in GS and NS of ZigBee represents end to end delay of all packets received by the 802.15.4 MACs of all WPAN nodes in the network and forwarded to the higher layer. MAD represents the total time of queuing and contention delays of the data frames transmitted by all the 802.15.4 MAC. For each frame, this delay calculated as the duration from the time when it is inserted into the transmission queue, which is arrival time for higher level data packets and creation time for all other frames types, until the time when the frame is sent to the physical layer for the first time.
\begin{table}
\caption {Traffic Parameters of ZigBee}
\begin {center}
\begin {tabular} {| p{3cm} | p{3cm} |}
\hline
Attribute & Value \\ \hline
Packet Interarrival Time & 0.01 bit/second \\ \hline
Packet Size & 1024 bytes \\ \hline
Start Time & uniform (Min 20,Max 21) \\ \hline
Stop Time & Infinity \\ \hline
\end{tabular}
\end{center}
\end{table}

\indent Fig 3 shows delay and MAD of ZigBee in GS. Time is placed on the $x$-axis and delay statistics are placed on the $y$-axis. Graph of both delay
and MAD increases with passage of time. As load is increasing, there is increase in the queuing size of ZigBee, due to which delay increases. ZigBee works on $CSMA/CA$ so there is collision, as load in network increased there is more collision and hence delay is large as compared to other devices.
\begin{figure}[!b]
\centering
\caption{DELAY AND MAD OF ZigBee In GS}
\includegraphics[width=3.5 in, height=2.5 in]{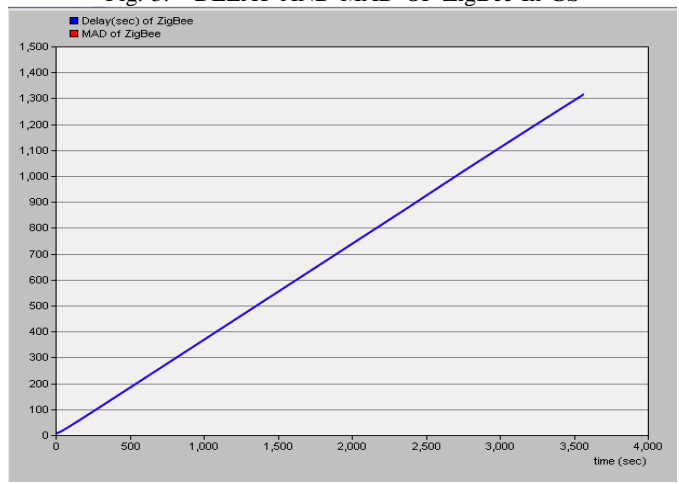}
\end{figure}

\indent Fig 4 shows the delay and MAD of WLAN in GS. Delay of WLAN increases with time due to increase in load and in amount of data. Delay also increases in WLAN due to the interference of ZigBee with it because ZigBee supports low data rate and WLAN have to wait for data to arrive and forward it to IP-Cloud. MAD of WLAN increases with time, however, as load increases, delay becomes constant. This is because as large data is passed in the start so WLAN experiences more MAD, and as load becomes constant MAD of WLAN also becomes constant.
\begin{figure}[!b]
\centering
\caption{DELAY AND MAD OF WLAN In GS}
\includegraphics[width=3.5 in, height=2.5 in]{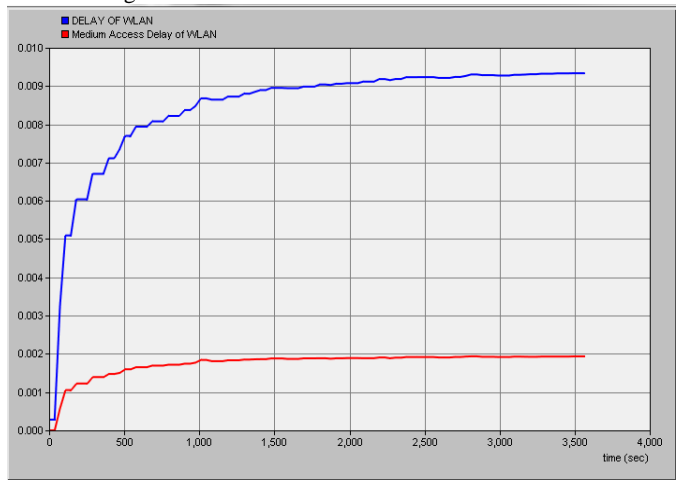}
\end{figure}
\\
\indent Fig 5 shows graphs of delay and MAD of ZigBee. Fig 5 tells about the behavior of delay and MAD of ZigBee in NS, delay and MAD are
very close to each other in this graph. Measurements of delay and MAD have been take at the router end, which is sending data to the WLAN. In the beginning
of simulations delay is minimum. As load increases, there is more queuing and collision hence, delay is increasing with time.
\begin{figure}[!b]
\centering
\caption{DELAY AND MAD OF ZigBee In NS}
\includegraphics[width=3.5 in, height=2.5 in]{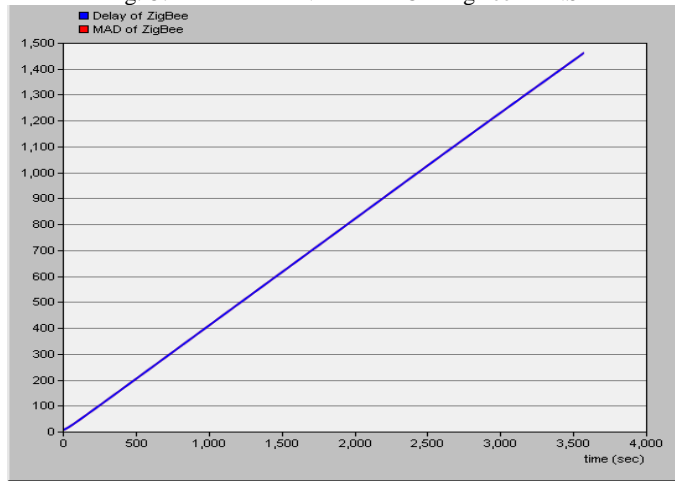}
\end{figure}
\\
\indent In fig 6 delay and MAD of WLAN in NS are shown. WLAN receives the data and sends it to its router for further process. Delay and MAD are calculated at the router end, which is communicating directly with the IP-cloud. Delay of WLAN decreases with time and after transmission of few data it becomes constant. As WLAN receiving data from ZigBee, its load is totally dependent on the traffic coming from ZigBee. As traffic becomes low, delay of WLAN becomes constant. Whereas MAD of WLAN increases time because, MAC layer has to wait for data for long time.
\begin{figure}[!h]
\centering
\caption{DELAY AND MAD OF WLAN In NS}
\includegraphics[width=3.5 in, height=2.5 in]{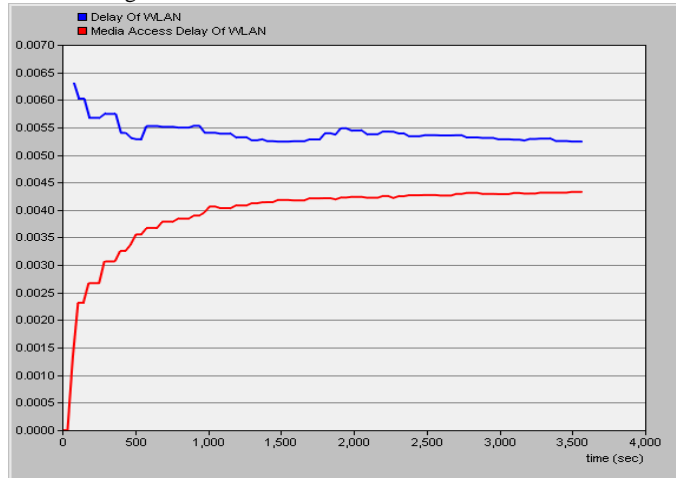}
\end{figure}
\subsection{Path 2}
\indent In this path data passes through ZigBee, WiMAX and IP-cloud before it reaches the monitoring room. Parameters are given in table IV. Destinations are set for each device, ZigBee End device destination is ZigBee coordinator and ZigBee coordinator destination is ZigBee router. Traffic attributes with their values are given in table VI. Attributes, whose graphs of the individual statistics have been plotted are as follow. Packet size, Interarrival time, start and stop time of this path are given in table VI.\\
\indent GS of ZigBee and WiMAX : Delay(sec)\\
\indent NS of ZigBee and WiMAX : Delay(sec)\\
\indent Delay in GS of WiMAX tells about end to end delay of all the packets received by the WiMAX MACs of all the WiMAX nodes in the network and forwarded to the higher layer. Delay of ZigBee tells about end to end delay of all packets received by the 802.15.4 from all PANs and forwarded it to the higher layers for further processing.\\
\indent Delay in NS of ZigBee represents the end to end delay of all packets received by the 802.15.4 MACs from each PANs node and forwarded it to the higher layer. In WiMAX, delay represents end to end delay of all packets received by the WiMAX MACs of each WiMAX node in the network and forwarded it to higher layer.
Fig 7, 8 and 9 shows the delay of ZigBee and WiMAX in GS and NS.\\
\indent In fig 7, delay of ZigBee is shown In GS. Delay is plotted as a function of time. With the passage of time, delay is increasing. ZigBee works on CSMA/CA mechanism due to which collision occurs as a result there is continuously increase in delay. This delay is also due to considerable increase of load, which is delivered to ZigBee. As more and more data is coming towards the MAC layer of ZigBee there will be more collision.
\begin{figure}[!b]
\centering
\caption{DELAY of ZigBee In GS}
\includegraphics[width=3.5 in, height=2.5 in]{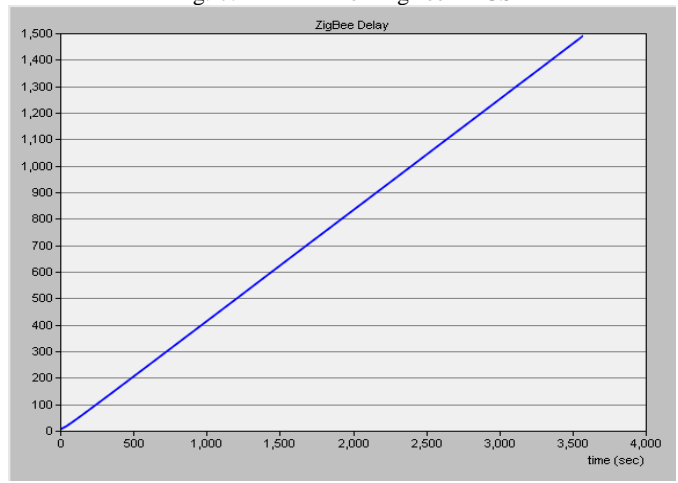}
\end{figure}
\\
\indent Fig 8 shows the delay of WiMAX plotted as function of time. WiMAX supports high data rates and have high capability of handling load due to which delay in WiMAX remains very low up to 0.005 and remains same throughout the time. ZigBee offers low data rates, so low data traffic is coming towards WiMAX hence, it not affects delay of WiMAX due to ability of performing highly at low data rate.
\begin{figure}[!b]
\centering
\caption{DELAY of WiMAX In GS}
\includegraphics[width=3.5 in, height=2.5 in]{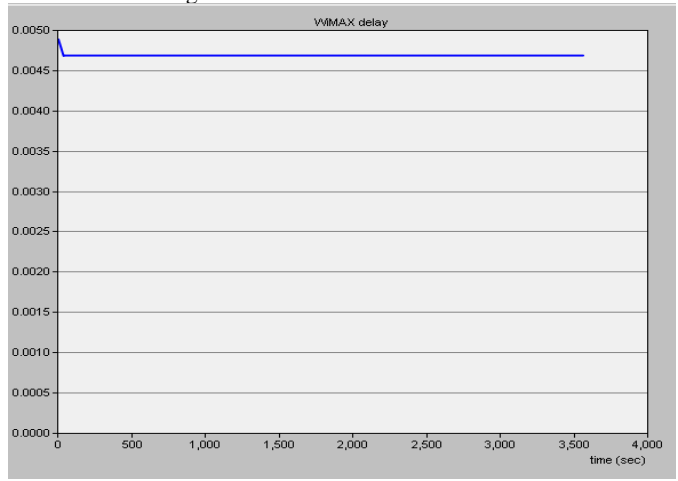}
\end{figure}
\\
\indent Fig 9 shows Delay of ZigBee in NS. This delay is quiet similar to ZigBee, reason of delay in GS is that, nodes are experiencing collision. As back off exponent is enabled therefore node experiencing collision goes to backoff state, where it senses the medium at random time and when it finds it free, it again sends the data. Minimum backoff exponent is kept at 2 and maximum is 3. With the passage of time delay in ZigBee is continuously increasing due to its low bandwidth and ability to support low data rates.
\begin{figure}[!h]
\centering
\caption{DELAY of ZigBee In NS}
\includegraphics[width=3.5 in, height=2.5 in]{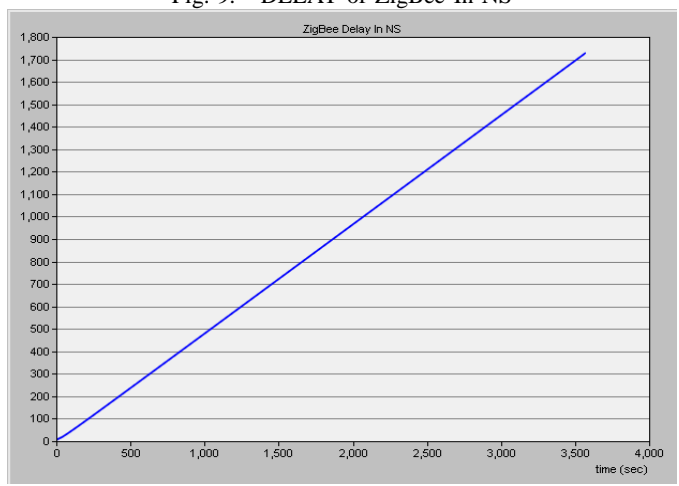}
\end{figure}
\subsection{Path 3}
Data from nodes passes through ZigBee, UMTS, IP-cloud and reaches health monitoring room. Parameters of this path are given in table V. Simulations have been run according to values given in table VI. Statistical attributes whose graphs are plotted as follows \\
\indent GS of ZigBee and UMTS: End to End delay(d) (Sec), Packet Delay (sec)\\
\indent NS of UMTS: End to End delay, Access Delay
\\
\indent End to End delay of UMTS in GS is also called as mouth to ear delay or analog to analog delay. This delay is given by equation 12.
\begin{eqnarray}
d&=&network_{delay}+encoding_{delay}+decoding_{delay} \nonumber \\
& & +compression_{delay} + decompression_{delay}
\end{eqnarray}
\indent In equation 12 network delay is, the delay of packet given to next component for encoding. Encoding and decoding delays are compute from encoding and decoding schemes. Compression and decompression delay comes from packet compression and decompression for further processing. Packet delay of UMTS in GS represents various end to end delay of various packets received by UMTS. \\
\indent End to End delay of UMTS in NS tells about the delay that UMTS experiences after data from ZigBee comes to user equipment and it passes it on to Node B of UMTS. Access Delay of UMTS in NS represents the delay that node experiences while accessing the network for data. It is the delay that is experience by node while contention procedure until data is send to next part for transmission. Fig 10, 11 and 12 shows the delays of ZigBee and UMTS in GS and NS.

Fig 10 shows the packet and end to end delay of UMTS in GS. Both delays are decreasing with passage of time, due to decrease in amount of data that has been sent to UMTS by ZigBee. After all of data that has been send from nodes to ZigBee is completed then delay in UMTS is constant. UMTS is totally a cellular based structure so it experiences hand over problems. Delay in UMTS occurs due to hand over problems that affects performance of UMTS.
\begin{figure}[!t]
\centering
\caption{Packet and End to End DELAY of UMTS In GS}
\includegraphics[width=3.5 in, height=2.5 in]{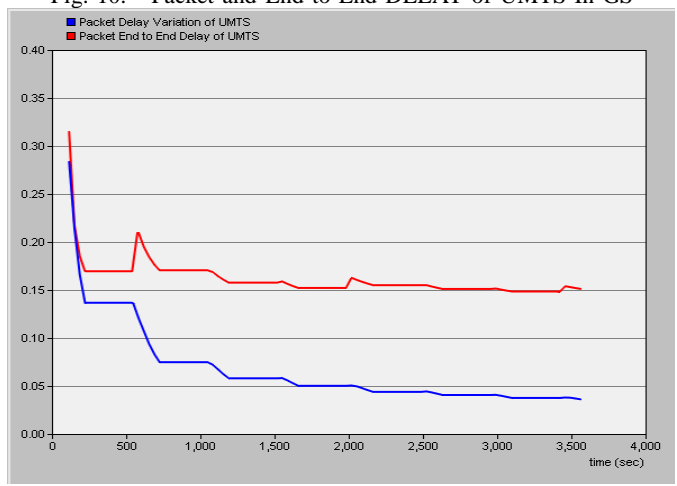}
\end{figure}

Fig 11 shows the packets and end to end delay of ZigBee in GS. Both delays are increasing linearly like in case of ZigBee connected to WiMAX. Delay in ZigBee occurs due to collision, as number of queuing of packets increases, there is more delay, with nodes going to backoff state and continuously sensing medium to send data, therefore results in more collision. Due to low data rates of ZigBee, it is unable for it to send data at high speed to UMTS.
\begin{figure}[!t]
\centering
\caption{Packet and End to End Delay of ZigBee in GS}
\includegraphics[width=3.5 in, height=2.5 in]{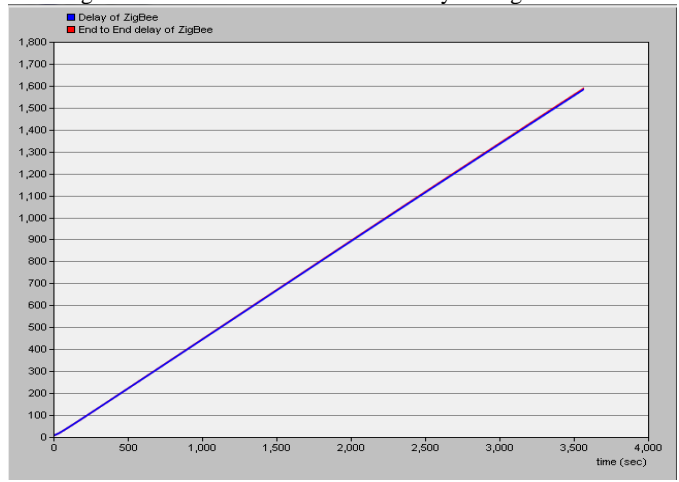}
\end{figure}
\begin{figure}[!t]
\centering
\caption{Access and End to End Delay of UMTS in NS}
\includegraphics[width=3.5 in, height=2.5 in]{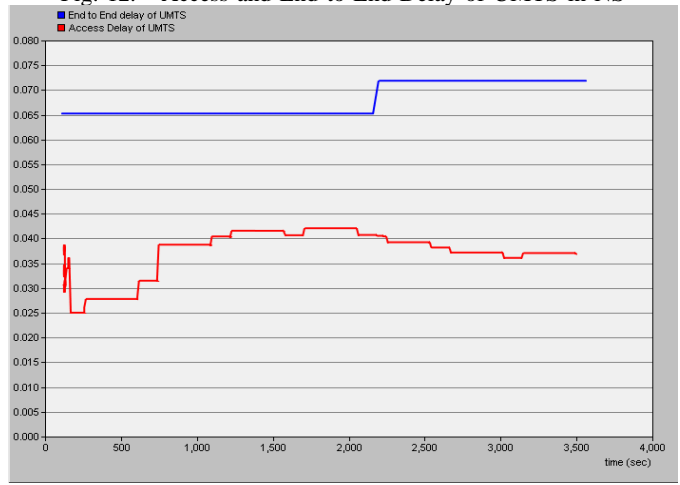}
\end{figure}
\\
\indent Fig 12 shows the end to end and packet variation delay of UMTS in GS. End to End delay remains constant for some time then there is a sudden increase in the delay this is because change of the the parts through which data is going in UMTS. Delay in user end which receives data remains constant, as data is send to other part of UMTS for further processing, there is delay due to sudden sending of data. Access delay in the beginning of simulations rises then there is a small dip. After some time there is again a rise in the access delay of UMTS in NS.
\section{Conclusion and Future Work}
In this paper, delay performance of transmitting medical data from sensors connected to body or implanted, over a heterogeneous multi-hop environment through three different paths have been analyzed. Wireless transmission of medical data goes through different networks in each path, networks are established on the basis of environmental conditions, structure of network and availability of device setup. ZigBee is common in each path which is connected, it receives data from sensor nodes and passes it to network connected. There are three paths available WLAN, WiMAX and UMTS connected with ZigBee. Delay of each path is calculated and simulated. Based on our analysis, overall transmission delay of each path can be evaluated over three hop wireless channel. This work focuses on calculation of delay of each link connected in each path. Future work includes calculation of overall transmission delay, sending of data from nodes and receiving instruction from health monitoring room through different paths.


\begin{thebibliography}{0}

\bibitem{1} M.A. Ameen, N. Ullah, and K. Kwak. Design and analysis of a mac protocol for wireless body area network using wakeup radio. In
Communications and Information Technologies (ISCIT), 2011 11th International Symposium on, pages 148–153. IEEE, 2011.

\bibitem{2} S.J. Hsu, H.H. Wu, S.W. Chen, T.C. Liu, W.T. Huang, Y.J. Chang, C.H. Chen, and Y.Y. Chen. Development of telemedicine and
telecare over wireless sensor network. In Multimedia and Ubiquitous Engineering, 2008. MUE 2008. International Conference on,
pages 597–604. IEEE, 2008.

\bibitem{3} J.Y. Khan, M.R. Yuce, and F. Karami. Performance evaluation of a wireless body area sensor network for remote patient monitoring. In
Engineering in Medicine and Biology Society, 2008. EMBS 2008. 30th Annual International Conference of the IEEE, pages 1266–1269.
IEEE, 2008.
\bibitem{4} B. Latre, B. Braem, I. Moerman, C. Blondia, E. Reusens, W. Joseph, and P. Demeester. A low-delay protocol for multihop wireless
body area networks. In Mobile and Ubiquitous Systems: Networking \& Services, 2007. MobiQuitous 2007. Fourth Annual International
Conference on, pages 1–8. IEEE, 2007.
\bibitem{5} B. Latr´e, P.D. Mil, I. Moerman, B. Dhoedt, P. Demeester, and N.V. Dierdonck. Throughput and delay analysis of unslotted ieee 802.15.
4. Journal of Networks, 1(1):20–28, 2006.
\bibitem{6} X. Liang and I. Balasingham. Performance analysis of the ieee 802.15. 4 based ecg monitoring network. Proceedings of the 7th
IASTED International Conferences on Wireless and Optical Communications, pages 99–104, 2007.
\bibitem{7} B. Sidhu, H. Singh, and A. Chhabra. Emerging wireless standards-wifi, zigbee and wimax. World Academy of Science, Engineering
and Technology, 25:1345–1349, 2007.
\bibitem{8} W. Song. Heterogeneous multi-hop transmission of compressed ecg data from wireless body area network. In Communications (ICC),
2011 IEEE International Conference on, pages 1–5. IEEE, 2011.
\bibitem{9} S. Ullah and K. Kwak. An ultra low-power and traffic-adaptive medium access control protocol for wireless body area network. J.
Med. Syst, 2010.
\bibitem{10} M.K. Yoon, J.E. Kim, K. Kang, K.J. Park, M.Y. Nam, and L. Sha. End-to-end delay analysis of wireless ecg over cellular networks.
In Proceedings of the 1st ACM international workshop on Medical-grade wireless networks, pages 21–26. ACM, 2009.
\end{thebibliography}
\end{document}